\documentclass[10pt, conference, compsocconf]{IEEEtran}
\usepackage{subcaption}
\usepackage{pgfplots}
\pgfplotsset{compat=1.3}
\usepackage{adjustbox}
\usepackage{listings}
\usepackage{graphicx}
\usepackage{enumitem}
\usepackage{cleveref}
\usepackage{booktabs}
\usepackage{omplst}
\newcommand{\highlightForReview}[1]{\textcolor{black}{#1}}

\begin{document}
%
\title{Worksharing Tasks: an Efficient Way to Exploit Irregular and 
Fine-Grained Loop Parallelism}

\author{
\IEEEauthorblockN{1\textsuperscript{st} Marcos Maro\~nas}
\IEEEauthorblockA{\textit{Barcelona Supercomputing Center (BSC)}\\
marcos.maronas@bsc.es}
\and
\IEEEauthorblockN{2\textsuperscript{nd} Kevin Sala}
\IEEEauthorblockA{\textit{Barcelona Supercomputing Center (BSC)}\\
kevin.sala@bsc.es}
\and
\IEEEauthorblockN{3\textsuperscript{rd} Sergi Mateo}
\IEEEauthorblockA{\textit{Barcelona Supercomputing Center (BSC)}\\
sergi.mateo@bsc.es}
\and
\IEEEauthorblockN{4\textsuperscript{th} Eduard Ayguad\'{e}}
\IEEEauthorblockA{\textit{Barcelona Supercomputing Center (BSC)}\\
\textit{Universitat Polit$\grave{e}$cnica de Catalunya (UPC)}\\
eduard.ayguade@bsc.es}
\and
\IEEEauthorblockN{5\textsuperscript{th} Vicen\c{c} Beltran}
\IEEEauthorblockA{\textit{Barcelona Supercomputing Center (BSC)}\\
vicenc.beltran@bsc.es}
}

%
%
%
%


\maketitle

\begin{abstract}

Shared memory programming models usually provide worksharing and task 
constructs. The former relies on the efficient fork-join execution model to 
exploit structured parallelism; while the latter relies on fine-grained 
synchronization among tasks and a flexible data-flow execution model to exploit 
dynamic, irregular, and nested parallelism. On applications that show both 
structured and unstructured parallelism, both worksharing and task constructs can 
be combined. However, it is difficult to mix both execution models without 
penalizing the data-flow execution model. Hence, on many applications structured 
parallelism is also exploited using tasks to leverage the full benefits of a 
pure data-flow execution model. However, task creation and management might 
introduce a non-negligible overhead that prevents the efficient exploitation of 
fine-grained structured parallelism, especially on many-core processors.
In this work, we propose \textit{worksharing tasks}. These are tasks that 
internally leverage worksharing techniques to exploit fine-grained structured 
loop-based parallelism. The evaluation shows promising results on several 
benchmarks and platforms. 

\end{abstract}


\section{Introduction}
\label{sec:intro}
\vspace{-0.1cm}
The introduction of the first multiprocessor architectures led to the 
development of shared memory programming models. One of those is OpenMP, which 
became a \textit{de facto} standard for parallelization on shared memory 
environments.

OpenMP~\cite{openmp}, with its highly optimized fork-join execution model, is a 
good choice to exploit structured parallelism, especially when the number of 
cores is small. Worksharing constructs, like the well-known \texttt{omp for} 
construct, are good examples of how OpenMP can efficiently exploit structured 
parallelism. However, when the number of cores increase and the work 
distribution is not perfectly balanced, the rigid fork-join execution model can 
hinder performance.

The \texttt{omp for} construct accepts different scheduling policies that can
mitigate load-balancing issues; and the \texttt{nowait} clause avoids the 
implicit barrier at the end of an \texttt{omp for}. Still, both techniques are 
only effective in a few specific situations. Moreover, the fork-join execution 
model is not well-suited for exploiting irregular, dynamic, or nested 
parallelism.

Task-based programming models were developed to overcome some of the
above-mentioned limitations. The first tasking models were based solely on the
tasks and taskwaits primitives, which naturally support irregular, dynamic, and
nested parallelism. However, these tasking models are still based on the 
fork-join execution model. The big step forward came with the introduction of 
data dependences. Thus, replacing the rigid fork-join execution model by a more
flexible data-flow execution model that relies on fine-grained synchronizations
among tasks. Modern task-based programming models \highlightForReview{such as
as Cilk, OmpSs or OpenMP tasking model} have evolved with advanced features to 
exploit nested parallelism~\cite{weakdeps}, hardware accelerators 
\cite{ayguade2009proposal}\cite{duran2011ompss}\cite{augonnet2011starpu}, and
seamless integration with message passing APIs such as MPI~\cite{Sala2018b}. 

The flexibility of the data-flow execution model relies on the dynamic management
of data-dependences among tasks. However, dependences management might introduce 
a non-negligible overhead depending on the granularity and number of tasks. 
Hence, finding the adequate task granularity becomes a key point to get good 
performance: too many fine-grained tasks will increase task overheads, but too 
few coarse-grained tasks will hinder the available parallelism. Yet, it is not 
always possible to reach the optimal granularity that is coarse enough to
compensate for the overheads while opening sufficient parallelism. Moreover, the 
granularity is limited by the problem size per core. Thus, if the problem size 
per core is too small, the granularity might be suboptimal, hurting the 
performance.

For those situations, it makes sense to combine both strategies---tasking and
worksharing---in a way that we can palliate the drawbacks of each strategy
while maximizing their strengths. To do so, we propose an extension to tasks 
especially adapted to avoid the fork-join execution model. We introduce the 
\texttt{for} clause applied to the task construct in the OmpSs-2 programming 
model \cite{ompss2}.



\highlightForReview{Our evaluation shows promising results.} The \texttt{task 
for} construct enlarges the set of granularities where peak performance is 
reached. Additionally, it enables users to increase peak performance by up to 
10x compared to pure task-based implementation for small problem sizes with low 
parallelism.

To sum up, the contributions of this paper are (1) a proposal and 
implementation of worksharing tasks, which is a mechanism combining tasking and 
worksharing parallelization strategies; and (2) an evaluation of such anto 
implementation on several benchmarks and two different architectures.

The rest of this document is structured as follows: Section~\ref{sec:background} 
contains the motivations behind this work; Section~\ref{sec:related} reviews the 
most relevant related work; Section~\ref{sec:design} explains the changes 
affecting the model; Section~\ref{sec:implementation} details our 
implementation; Section~\ref{sec:evaluation} consists of an evaluation and 
discussion of the proposal; Section~\ref{sec:conclusions} summarizes the work 
done and provides concluding remarks; and, finally, Section~\ref{sec:future} 
presents future work proposals.

\section{Motivation}
\label{sec:background}

Fine-grained loop parallelism can be found in most HPC applications.
So, it is important to develop techniques that perform well for 
this kind of parallelism. Harris et al.~\cite{harris2015callisto}, already  
explored the importance of properly supporting fine-grained loop parallelism.

\highlightForReview{Nowadays, developers can use loop-based parallelism or task-based parallelism 
for coding their applications containing fine-grained loop parallelism.} Loop-based 
parallelism is quite simple to write, and it performs well in architectures with 
a low number of cores and applications with a small load imbalance. Despite this, 
it implies a rigid synchronization resulting in performance drops when 
facing many-core architectures and imbalanced applications. Task-based 
parallelism allows a data-flow execution, which is more flexible than its 
loop-based counterpart. Additionally, it provides several key benefits, 
previously mentioned in Section~\ref{sec:intro}. Thus, it usually performs well 
in many-core architectures and load imbalanced applications. 

Still, \highlightForReview{an inherent problem of task programming is 
granularity choice.} If task granularity is not adequately set, overhead may 
penalize overall performance. The overhead of tasks is caused by several 
different sources. The first one is the actual task creation, which usually 
implies costly dynamic memory allocations. Secondly, the computation of 
dependences between tasks, which involves the use of dynamic and irregular 
data-structures. Finally, the scheduling of the tasks across many cores can also 
become a bottleneck.

Task granularity and the number of created tasks are inversely proportional. 
\highlightForReview{Consequently, a given problem can be solved either by using 
many fine-grained tasks or a few coarse-grained ones.} Thus, finding an adequate 
granularity is a key point to optimally exploit resources when using 
tasks~\cite{navarro2017adaptive}, alleviating the aforementioned overheads, but 
still creating enough parallelism to maximize resource utilization. 

\begin{figure}
\label{fig:overhead}
\includegraphics[width=\columnwidth,height=4cm]{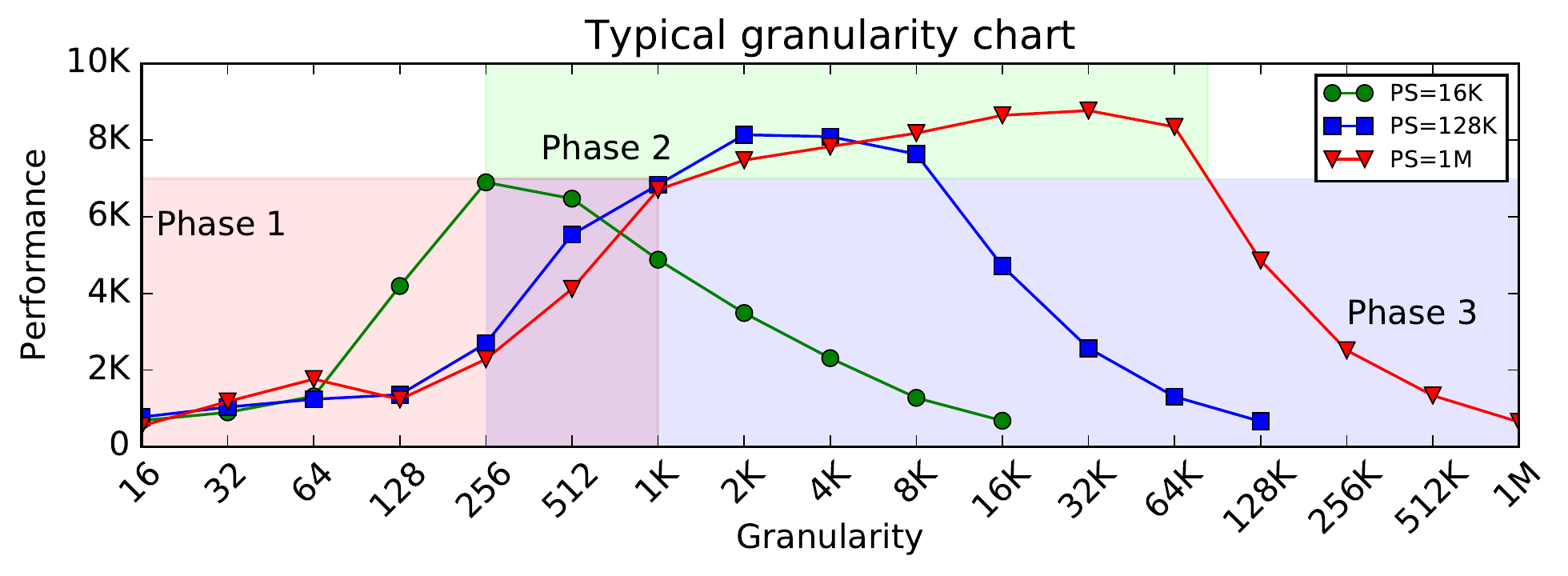}
\vspace{-0.70cm}
\caption{Typical granularity chart.}
\vspace{-0.5cm}
\label{fig:granularity}
\end{figure}

A typical granularity chart is shown in Figure~\ref{fig:granularity}. The x-axis 
varies the granularity of tasks, while the y-axis represents performance. 
\highlightForReview{The chart presents the results of the synthetic benchmark 
shown in Code~\ref{lst:synthetic}. There are three different series representing 
different problem sizes. The chart also contains coloured parts which represent 
different chart (not application) phases. Note that from X=256 to X=1K, phases 
1 and 3 are merged. This is because it is Phase 1 for PS=16K, but Phase 3 for 
PS=128K and PS=1M.}

When the problem size is 1M and 128K, there are three well-differentiated phases. 
In the first phase, we can see how the performance is low because there are too 
many very fine-grained tasks and the overheads of creation and management of 
that amount of small tasks are too costly. In the second phase, performance 
grows until reaching peak performance. Finally, in the third phase, performance 
decreases again because there is not enough parallelism (i.e., not enough tasks 
are being created to feed all the CPUs).

\begin{lstlisting}[caption=OMP\_F,label={lst:synthetic},escapechar=$]
for(size_t block = 0; block < NUM_BLOCKS; block++) {
    size_t start = block*TSIZE;
    size_t size = start+TSIZE > N ? N-start : TSIZE;
    #pragma oss task inout(a[start;size]) priority(block)
    for(size_t j2=start; j2 < start+size; j2++) {
        a[j2] += b[j2]*c[j2];
    }
}
\end{lstlisting}

\highlightForReview{
Typically, a good granularity allows having, at least, one work unit per core to 
occupy all the resources. Ideally, having more than one work unit per core is 
better to prevent some load balance problems. Additionally, it is important 
setting a granularity coarse enough to alleviate task management overheads. 
However, there is a crucial factor that can limit the granularity choice: the 
problem size per core.} The problem size per core is the result of dividing the 
total problem size by the number of available cores. In consequence, the problem 
size per core only depends on the total problem size and the number of cores 
available in our system.

In an ideal case, like the problem sizes of 1M and 128K in 
Figure~\ref{fig:granularity}, the granularity can grow until the overhead is not 
a problem. At that point, the problem size per core is big enough to create 
sufficient tasks---of a granularity that is not affected by the overhead---to 
feed all the resources. This happens in the second phase when peak performance 
is reached.

In contrast, if the problem size per core is not big enough, the developer must 
decide between a finer granularity that is still affected by the overhead but 
creates sufficient parallelism, or a coarser-granularity that is less affected 
by the overhead but causes a lack of parallelism. When this happens, the second 
phase of the typical granularity chart does not appear, being unable to reach 
peak performance. This phenomenon occurs in Figure~\ref{fig:granularity} when 
the problem size is 16K.

Daily, developers are involved in situations where the problem size per core 
is not optimal, jeopardizing the use of tasks in their applications.  
%
\subsubsection{\textbf{Strong scaling in distributed environments}}
This is a common scenario in HPC environments. Strong scaling starts from a 
given problem size per core, and make it smaller either by augmenting the 
number of resources or by decreasing the total problem size. As we have seen, 
reducing the problem size per core while maintaining the granularity of the 
tasks can lead to insufficient work.

\subsubsection{\textbf{Many-core architectures}}
Increasingly, architectures have more and more cores. This trend directly 
affects the problem size per core, which becomes reduced because the same 
problem size is divided among more resources. Thus, setting an adequate 
granularity becomes harder or even impossible, leading us to either increased 
overhead or lack of parallelism.

\subsubsection{\textbf{Applications requiring different granularities}}
Many applications rely on different kernels to perform a computation, and each 
of them may require a different task granularity to achieve optimal performance. 
Finding an adequate granularity that fits all the different algorithms may be 
impossible. For this case, it is especially important to have a broad set of 
granularities where peak, or at least acceptable, performance is reached because 
if all the kernels have several granularities that reach peak performance, it is 
easier to find a granularity that performs well for all the kernels than it 
would be if there is a single granularity getting peak performance for each 
kernel.

Additionally, it may happen that an application with different kernels must 
share the same granularity. The reason is that the data partitioning may 
implicitly set the task granularity. When this happens, it is especially 
important having a wide set of granularities performing well in all the kernels. 
This way, it is easier to find a coincidence across all the sets.

\highlightForReview{
\\Apart from this, granularity issues may prevent runtime libraries from 
developing sophitiscated and smart policies. Those policies may introduce few 
overhead per task but could provide benefits in terms of programmability and 
performance. However, if a program contains a huge number of tasks, the 
aforementioned small overhead per task, rapidly becomes unaffordable. A good 
example is the support for region dependences. This kind of dependences enables
users to annotate their codes with the whole memory regions a task actually 
access. Then, the runtime library computes the dependences with all the partial 
overlappings, actually preventing any task that shares even a single byte to 
execute until the current task finishes. In Code~\ref{lst:region_deps}, using 
region dependences, the second task depends on the first task, while it does not 
when using discrete dependences (e.g., OpenMP dependences) because those only 
consider the start address. Region dependences are very useful to simplify 
codes, but they come at a cost. The computation of the dependences is more 
expensive compared to discrete dependencies. In consequence, if the number of 
tasks is huge, the overhead may become excessive. 
}

\begin{lstlisting}[caption=Region deps,label={lst:region_deps},escapechar=$]
// a[0;8] means from 0 (included) to 8 (not included)
#pragma oss task inout(a[0;8])
task_body();

// a[2;6] means from 2 (included) to 6 (not included)
#pragma oss task inout(a[2;6])
task_body();
\end{lstlisting}

\highlightForReview{
}

\highlightForReview{
To sum up, task-based parallelism offers several key benefits that developers 
want to keep. Notwithstanding, there are currently several difficulties or 
problems when programming fine-grained loop parallelism with tasks. (1) 
Granularity is critical: for that purpose, a thorough and time-costly analysis 
must be done in order to choose it adequately; (2) adequate granularity does not 
always exist: some scenarios may force developers to choose either overhead or 
lack of parallelism; and (3) runtime libraries cannot develop sophisticated 
tasking management policies: those could jeopardize the performance in programs 
with a very large task number.
}



\section{Related work}
\label{sec:related}

Our proposal, based on the idea of hierarchical partitioning, is broadly used in 
distributed environments to reduce overheads. Most applications firstly 
partition data using inter-node parallelism, spreading such data among different 
nodes. Then, the work is partitioned again using intra-node parallelism. There 
are several works in the literature proposing several techniques based on this 
idea, such as \cite{dursun2009multilevel}, \cite{chatterjee2013integrating}, 
and \cite{rabenseifner2009hybrid}.

As OpenMP is the standard for shared memory parallelism, we performed a thorough 
review of the OpenMP environment to search related work. This can be seen in 
Section~\ref{subsec:openmp-related}. In addition, wider related work can be 
found in Section~\ref{subsec:non-openmp-related}.

\subsection{OpenMP related work}
\label{subsec:openmp-related}
On OpenMP, this hierarchical partition of data can be implemented in several 
ways using a combination of worksharing and tasking constructs.

One of those is using tasks to perform a first partition of the work, and then, 
each task contains a nested parallel region with a worksharing construct. The 
reason for using tasks in the first level of partitioning is the flexibility 
given by the data dependences. This implementation may increase the resource 
utilisation in some scenarios, boosting performance. However, we end up 
introducing a barrier inside each task, at the end of the nested parallel region. 

Barriers have been broadly treated in literature~\cite{nanjegowda2009scalability}.  
Currently, they are usually highly optimized so that they introduce only a few 
overhead in some situations, though, if the work is not perfectly balanced, the 
intrinsic rigidity of the fork-join model may lead to undesired waiting times. 
OpenMP introduced the \texttt{nowait} clause to palliate this issue. This clause 
omits the barrier at the end of a worksharing region. Nonetheless, this 
mechanism is not useful to avoid the barrier at the end of a parallel region 
inside a task. This is because the barrier is necessary to postpone the release 
of the task dependences until the work is completed.

OpenMP also provides different scheduling policies for the worksharing 
constructs, alleviating load imbalance problems. Still, they are not enough for 
many cases, and the rigidity of the fork-join model may lead to an 
underutilization of the resources.

It is possible to implement a different solution using task nesting. This is 
basically creating tasks inside tasks. So, a first partitioning is done using 
coarse-grained tasks with data dependences, which are then partitioned into 
fine-grained tasks without data dependences. The second level of partitioning 
reduces the overhead compared to a single level of partitioning where all the 
tasks have data dependences because the nested tasks do not pay the dependence 
management costs. In addition, using tasks improves load balance. 
However, tasking introduces other overheads associated with tasks management, 
such as dynamic allocations and task scheduling.

The OpenMP tasking model also provides the \texttt{taskloop} construct. Applied 
to a loop, it partitions the iteration space by generating several tasks. There 
is the possibility of specifying a \textit{grainsize} guaranteeing that each of 
the tasks created executes no less than \textit{grainsize} iterations. With this 
mechanism, the overhead may be reduced because allocations could be optimized to 
be done as a whole, instead of one by one. However, the number of tasks that
will be created and scheduled is still proportional to the problem size.

Task nesting could be done using the previously mentioned taskloop. Concretely, 
it is possible to implement a code that is firstly partitioned using tasks with 
dependences that contain taskloops with no dependences---in fact, taskloop 
does not support data dependences yet. Withal, this is very similar to the 
previously described implementation using pure task nesting.

\subsection{Non-OpenMP related work}
\label{subsec:non-openmp-related}
There exist other works, such as StarPU \textit{Parallel
tasks}~\cite{starpu-parallel-tasks}. A parallel task is a task that can be run 
in parallel by a set of CPUs, which might sound similar to our proposal of 
worksharing tasks. Nonetheless, these tasks are like the combination of an 
OpenMP task with a worksharing construct inside. Thus, it contains an implicit 
barrier at the end. Moreover, in many cases, only a single parallel task can run 
at a time. The reason for this is that many environments and libraries they 
leverage internally do not support concurrent calls to create several parallel 
regions without nesting. 

Intel Cilk presents the \texttt{cilk\_for}~\cite{cilk_for}, which is used to 
parallelize loops. The body of the loop is converted into a function that is 
called recursively using a divide and conquer strategy for achieving better 
performance. However, there is a \texttt{cilk\_sync} at the end of each 
iteration. Therefore, synchronization is quite rigid, similarly to OpenMP 
worksharings. 
\section{Worksharing tasks}
\label{sec:design}

In this section, we detail the semantics of our proposal, its constraints and 
how it fits into the tasking model. Also, we discuss its applicability and
utility.

\subsection{Syntax}
\label{subsec:syntax}
We propose a new clause for the task construct. This is the \texttt{for} clause 
for C/C++ and the \texttt{do} for Fortran. 

A \texttt{task for}---or worksharing task---accepts all the clauses accepted 
by a regular task except the final clause because \texttt{task for} is always 
final. Note that being final means that no tasks can be created inside the 
context of a worksharing task. Additionally, it accepts the  
\texttt{chunksize(integer-expr)} clause. The integer-expr specified as a 
chunksize sets the minimum chunk of iterations that each worker is going to 
execute when it requests work to the \texttt{worksharing} task, except for the 
last chunk that might contain fewer iterations. \highlightForReview{If not set, 
the default value is $Task size/NumberOfCollaborators$, so that each 
collaborator has at least one chunk to run.}

The \texttt{for} clause can only be applied to a task that is immediately 
succeeded by a loop statement. Codes \ref{lst:pipelining} 
and~\ref{lst:example_fortran} contain examples of code using the new clause.

\noindent\begin{minipage}{.55\columnwidth}

\begin{lstlisting}[caption={Code of Figure ~\ref{fig:pipelining}},
                   label={lst:pipelining},escapechar=$]
for(int i = 0; i < 2; i++) {
    #pragma oss task for [inout(a)]
    for(...){...}
    #pragma oss task for [inout(b)]
    for(...){...}
    #pragma oss task for [inout(c)]
    for(...){...}
}
\end{lstlisting}
\end{minipage}\hfill
\begin{minipage}{.35\columnwidth}
\begin{lstlisting}[caption={Fortran example},label={lst:example_fortran},escapechar=$]
!$\$$oss task do
do i=0, N
  call do_work();
end do
\end{lstlisting}
\end{minipage}
\vspace{-0.25cm}

\subsection{Semantics}
\label{subsec:semantics}
\begin{figure}
\centering
\includegraphics[width=\columnwidth,height=3.8cm]{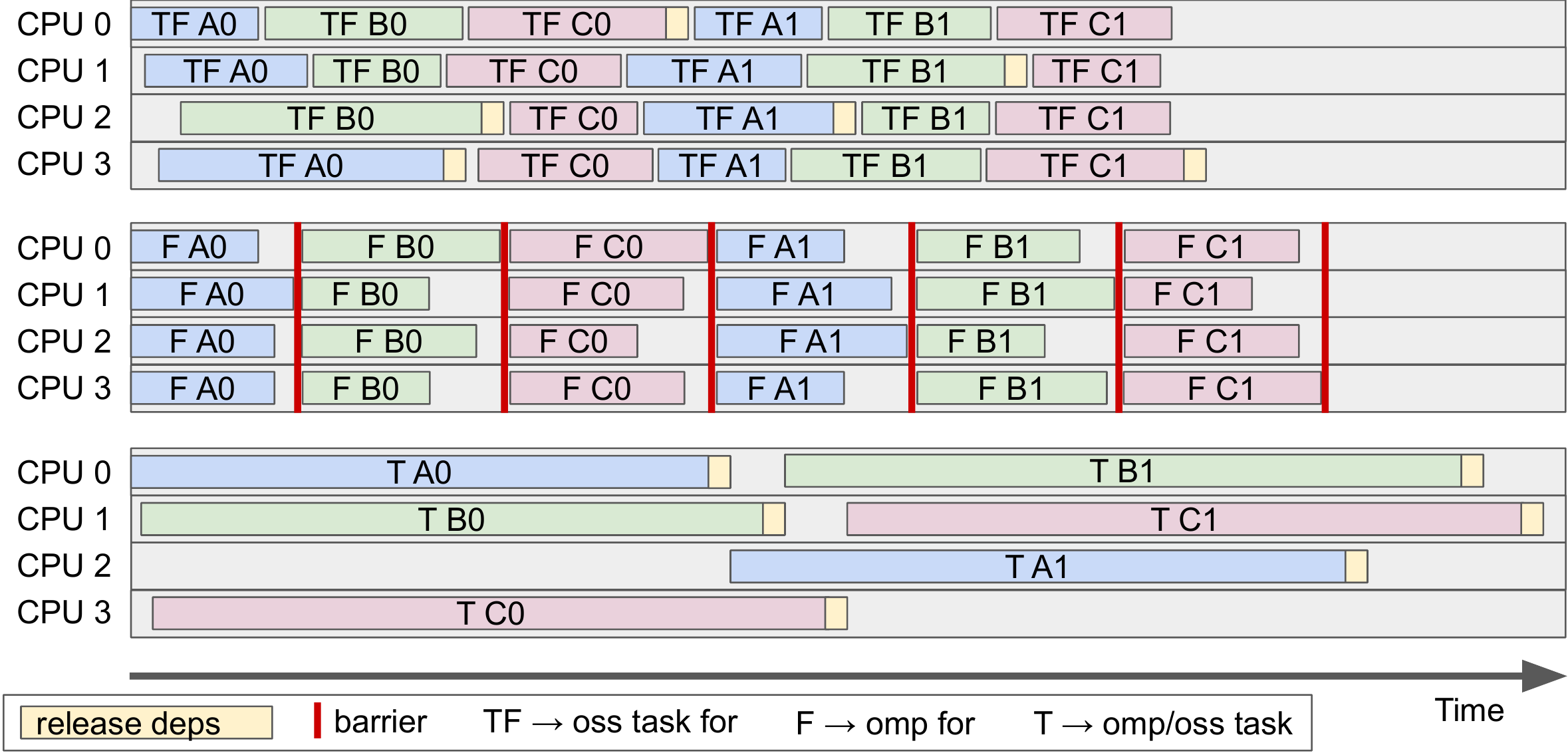}
\caption{Illustration of similar execution using OmpSs-2 worksharing tasks and 
    OpenMP worksharings.}
\vspace{-0.50cm}
\label{fig:pipelining}
\end{figure}

A \texttt{worksharing} task behaves like a regular task in almost everything. 
\highlightForReview{
The main difference is illustrated in Figure~\ref{fig:pipelining} whose code is 
shown in Code~\ref{lst:pipelining}.
}
Regular tasks are executed entirely by a single worker concurrently, while a 
\texttt{task for} may be executed collaboratively by several workers, as a 
worksharing construct. Nevertheless, one can see in Figure~\ref{fig:pipelining} 
that it does not imply any synchronization or barrier at all. A worksharing task 
is like a regular task in this sense, and the synchronization is done through 
data dependences or explicit synchronization points. Note that the data 
dependences of the worksharing tasks are released when the last chunk is 
finished by the thread that runs that last chunk. This can be seen in 
Figure~\ref{fig:pipelining}, represented by the small yellow piece at the end of 
the last chunk of each worksharing task.

As a worksharing construct, the iteration space of the for-loop is partitioned 
in chunks of \textit{chunksize} size. The key point is that these chunks do not 
have the usual overheads associated with a task---such as memory allocation and 
dependences management. To run a chunk, a thread only needs the boundaries of 
that chunk and the data environment, much like worksharing constructs. So, in 
summary, a worksharing task can be run in parallel by multiple threads, 
better amortizing the task management overheads. 

Usually, programmers use coarse granularities to overcome tasking overheads.
Using tasks, coarse granularities may limit parallelism, causing some resources 
to be idle as in the bottom part of Figure~\ref{fig:pipelining}. In contrast, 
using coarse-grained worksharing tasks, the work is split again into several 
fine-grained chunks that can be run concurrently by several workers. Thus, 
preventing resources from becoming idle and maximizing resource utilization, as 
shown in the top part of Figure~\ref{fig:pipelining}. 

Regarding chunk distribution, a worksharing task is highly flexible. The only 
guarantee is that work is partitioned in chunks of \textit{chunksize} 
size and it is executed at most by N collaborators of the same 
\textit{team}. 

A worksharing task creates a \textit{worksharing region} that is executed by a 
\textit{team} of workers. One important property of worksharing regions is 
illustrated in Figure~\ref{fig:pipelining}: up to N threads \textbf{may} 
collaborate on the completion of the work, but they are not forced to do so. 
This behavior happens with \textit{TF A0} and \textit{TF B0} which are run only 
by three threads while their team contains four threads. This happens because 
all the work has been assigned prior to the arrival of the last thread, so the 
last thread just goes ahead and gets more work.

A further key feature can be observed in Figure~\ref{fig:pipelining}. CPUs can 
leave the worksharing region before the actual completion of the whole 
worksharing task. CPU0 finishes its \textit{TF A0} chunk while CPU1 and CPU3 are 
still completing their chunks. However, instead of waiting as a regular 
worksharing does in the middle part of the figure, it moves forward to 
\textit{TF B0}. In other words, worksharing regions do not contain implicit 
barriers at the end. This behavior is equivalent to set a \texttt{nowait} clause 
in OpenMP worksharing constructs.

This feature is especially important because it permits the pipelining of 
different worksharing regions. This behavior can be observed in 
Figure~\ref{fig:pipelining}. For instance, when CPU1 finishes its \textit{TF A0} 
chunk, there is no remaining work in \textit{TF A0}. Hence, it leaves that 
worksharing region and joins \textit{TF B0}. However, \textit{TF A0} was still 
in execution by CPU3.

In summary, worksharing tasks implicitly alleviate the effects of a possible 
load imbalance through the ability of collaborators to leave a worksharing 
region when there is no remaining work. Thus, threads can just go forward and 
get more work instead of becoming idle waiting at a barrier. 
Worksharing tasks also palliate the granularity issues by allowing the use of 
coarse granularities that are partitioned anew at an additional level of 
parallelism. So, task management overheads are minimized, and resource 
utilization is maximized.

\subsection{Integration in OmpSs-2}
The concept of worksharing task is completely integrated into the model since 
at all levels it is like a task, except that it may be executed by several 
workers instead of by a single one. For that reason, it can interact with 
regular tasks without further problem using regular mechanisms: data dependences 
and explicit synchronization points regarding synchronization; and data-sharings 
for managing how the data is shared across different pieces of work. 

\subsection{Applicability}
\label{subsec:applicability}
\highlightForReview{
Worksharing tasks applicability is as wide as OpenMP worksharings. If the 
iterations in a loop are independent, then worksharing tasks can be applied.
} 
\highlightForReview{
Worksharing tasks are especially useful to deal with applications containing 
multiple kernels especially if those present different patterns 
(regular/irregular). Worksharing tasks enables users to program using a pure 
data-flow model while efficiently exploiting fine-grained loop parallelism. 
Some benchmarks that we expect to benefit from worksharing tasks and we are 
looking at, or plan to do so in the future are: Strassen, XSBench, SparseLU, 
MiniTri, Jacobi and miniAMR, to name a few. 
}

\subsection{Utility}
\label{subsec:utility}
\highlightForReview{
Worksharing tasks mitigate or solve the problems presented in 
Section~\ref{sec:background}. Firstly, worksharing tasks enlarge the set of 
granularities that deliver good performance. In scenarios where only a few tasks 
are created and if these are not enough to keep all the resources busy, the use 
of worksharing tasks mitigate the lack of parallelism. Thus, providing several 
extra granularities that still work well compared to regular tasks, overall, 
easing the granularity choice.
}

\highlightForReview{
Furthermore, as we already discussed in Section~\ref{sec:background}, there are 
scenarios when a good granularity does not exist and developers incur either on 
overhead or lack of parallelism. For that scenarios, worksharing tasks are 
especially useful because developers can reduce overhead by setting coarser 
granularities, without fearing a lack of parallelism. Given that worksharing 
tasks split the work among a whole team of collaborators, the total number of 
tasks required to keep all the resources busy is reduced from the total number 
of cores to the total number of teams. Hence, offering a solution to scenarios 
where tasks are unable to perform well.
}

\begin{figure}
\centering
\includegraphics[width=\columnwidth,height=3.8cm]{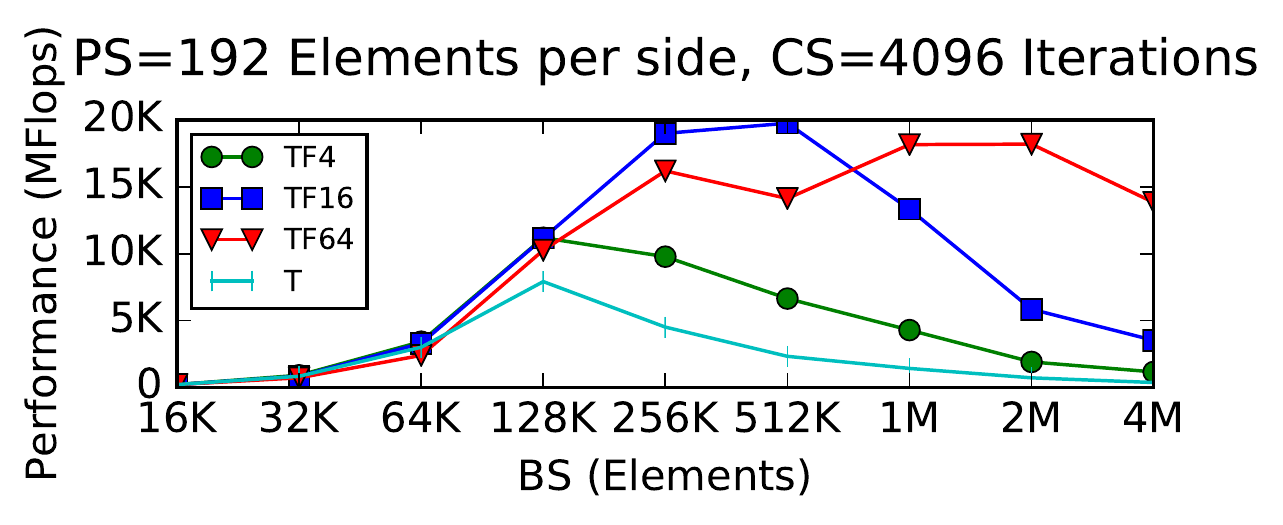}
\caption{Comparison of HPCCG benchmark using region dependences with regular 
    tasks and worksharing tasks.}
\vspace{-0.50cm}
\label{fig:regions}
\end{figure}

\highlightForReview{
Finally, since worksharing tasks are able to reduce the number of tasks 
by making them coarser without any significant performance loss, runtime 
libraries can develop sophisticated mechanisms to deal with task management.
One example can be seen in Figure~\ref{fig:regions}, where region dependences 
are in use. They are not suitable to be used with regular tasks given its low 
performance. However, they become suitable when combined with worksharing tasks.
}

\section{Implementation Details}
\label{sec:implementation}
We have implemented the concept of worksharing tasks in the OmpSs-2 programming 
model, which relies on the Mercurium source-to-source compiler~\cite{mercurium}
and the Nanos6 runtime library~\cite{nanos6}. In this section, we detail the 
extensions performed in both components to support worksharing tasks.

\subsection{Mercurium compiler}
We have extended the Mercurium compiler to support the new \texttt{for} clause 
applied to the \texttt{task} construct. Though, as this combination of 
\texttt{task for} can only be applied to loop statements, Mercurium is also in 
charge of checking so. In the same line, Mercurium also checks that final clause 
is not applied to a \texttt{task for} since it is not valid. 

Given that a worksharing task may be executed by several different threads, 
each of them should have a correct data environment to avoid possible errors in 
the computations. Sometimes, this may imply firstprivate or private 
data-sharings. Mercurium has the responsibility of providing a valid data 
duplication method to the runtime if firstprivate or private data-sharings have 
been set by the user. Otherwise, the runtime may not know how to duplicate data.

\subsection{Nanos6 runtime library}
Regarding the runtime library, firstly, we have extended the work descriptor of 
a task to include some extra information that permits splitting and spreading of 
the work. Basically, we add information about the boundaries of the loop and the
chunksize. This information is taken at task creation time. 

Then, when the worksharing task becomes ready, it is enqueued as the 
rest of the tasks. Eventually, the task is scheduled for execution. At this 
point, worksharing tasks follow a different path from regular tasks. Regular 
tasks are assigned to a worker thread, and it is in charge of executing the 
task and release its dependences if any. Worksharing tasks are also initially 
assigned to a worker thread but, instead of executing the whole task itself, it 
shares the task with its team.

Currently, in our implementation, the maximum size of the teams is defined at 
the beginning of the execution and remain unchanged until the end. 
Moreover, all the teams have the same maximum size. We build teams by grouping 
together hardware threads that are close in the physical layout.

The way a worksharing task is actually executed also differs from regular tasks. 
While for regular tasks we simply assign a work descriptor with its respective 
data environment to a thread and it just runs; worksharing tasks need some 
further steps. First of all, as several workers may collaborate to do the work, 
each of them needs its own work descriptor and data environment to avoid 
interferences caused by the concurrency. 

Each CPU has a preallocated extended work descriptor. When this CPU receives 
a chunk, the preallocated extended work descriptor is filled with the actual 
information of the chunk that has been assigned. This represents the control 
information for running its part of the worksharing task.

Regarding the data environment, given that we do not know in advance 
how many collaborators will be---neither which of them---, each worker triggers 
the duplication of the data environment in a lazy way. The worker triggers 
the data duplication when it has received work to do, and it has filled its 
preallocated work descriptor with the control information. Using the data 
duplication method provided by the compiler, the runtime duplicates the data 
environment and assigns it to the preallocated work descriptor of the thread.

Once a CPU already has the work descriptor and the data environment, it can 
start running its part of the work. The assignment of work from a worksharing 
task is done on a first-come-first-serve basis. It is guaranteed that a worker 
never receives fewer iterations than those specified in the chunksize clause, 
except the last chunk if there are not enough iterations to complete a chunk. 
However, it may receive several chunks. 

In the current implementation, the chunk scheduling policy is very similar to 
the \texttt{guided} scheduling policy of OpenMP since the number of assigned 
chunks is proportional to the number of unassigned chunks divided by the number 
of collaborators. Note that no matter how many chunks a collaborator receives, 
it performs the work descriptor filling and the data environment duplication 
only once per work request. After terminating the assigned chunks, a thread 
checks if it is the last. If so, the worksharing task has finished all the 
chunks, and as a result, it has finished as a whole. Data dependences, if any, 
are released at this moment. 

Otherwise, when a worker finishes its assigned chunks but the whole worksharing 
task has not finished there exist two possibilities: (1) all the work has been 
assigned and other collaborators are still running; (2) there is still work to 
be assigned. In (1), the worker that finishes its chunks just leaves the team 
and tries to get new work. In (2), the worker requests more chunks to the
current worksharing task.

It is also important to highlight that assigning chunks to a worker and 
finishing those chunks imply some overheads that regular tasks do not have. 
Even though we have tuned our implementation to allow fine-grained chunks, 
setting an adequate chunksize is important for the proper exploitation of 
worksharing tasks as shown in Section~\ref{subsec:cs-granularity}. Furthermore, 
the process of requesting work crosses the scheduler path. So, it has some 
associated locks that may be taken into account when setting the chunksize.
  
\section{Evaluation}
\label{sec:evaluation}
In this section, we provide an evaluation of our proposal, as well as a 
discussion of the results. First of all, we introduce the environments and 
platforms in which the experiments were conducted. Following, the benchmarks 
used and the different implementations developed are described. Then, for each 
experiment, we detail the methodology followed along with the results and 
discussion about them.

We include three different experiments. The first one is a granularity analysis 
on a many-core system. The objective is to show how the traditional ways of 
exploiting parallelism may easily suffer from a lack of parallelism on 
many-core architectures. The second experiment is a chunksize granularity 
analysis which aims to stress the importance of an adequate chunksize. 
Finally, the third experiment is a strong scaling experiment to illustrate some 
scenarios where the problem size per core prevents setting a good task 
granularity.

We wish to remark that OmpSs-2 implements region dependences. In contrast, 
OpenMP implements discrete dependences. Given that difference, we have 
introduced a new dependency system that supports discrete dependences in OmpSs-2
to make comparisons fairer.

\vspace{-0.10cm}
\subsection{Environment}
\label{subsec:env}

The experiments were carried out on two different platforms. The first platform 
is composed of nodes with 2 sockets Intel Xeon Platinum 8160 2.1GHz 24-core and 
96GB of main memory. The second platform is composed of nodes 
with 1 socket Intel Xeon Phi CPU 7230 1.3GHz 64-core and 96GB of main memory 
plus 16GB of high bandwdith memory.

Regarding the software, we used the Mercurium compiler (v2.3.0), the Nanos6 
runtime library, the gcc and gfortran compilers (v7.2.0), and the Intel 
compilers (v17.0.4).

\highlightForReview{
We would like to highlight that all the experiments have been run using the 
interleaving policy offered by the numactl command, spreading the data evenly 
across all the available NUMA nodes, in order to minimize the NUMA effect.
}

\subsection{Benchmarks}
\label{subsec:benchmarks}
We have considered four different benchmarks for the evaluation: the High 
Performance Computing Conjugate Gradient (HPCCG)~\cite{hpccg}, the matrix 
multiplication kernel (MATMUL), the N-body simulation and the Stream 
benchmark~\cite{stream}. HPCCG and the Stream benchmark were selected 
as representants of memory-bounded benchmarks while MATMUL and the N-body 
simulation represent computed-bounded workload.

\noindent\begin{minipage}{.45\columnwidth}
\begin{lstlisting}[caption=OMP\_F(S/D/G),label={lst:omp_for},escapechar=$]
#pragma omp for \
schedule( \
[static/dynamic/guided],TS)
for(i=0; i<PS; i++)
  do_work(i);
\end{lstlisting}
\end{minipage}\hfill
\begin{minipage}{.45\columnwidth}
\begin{lstlisting}[caption=OMP\_T/OSS\_T,label={lst:task},escapechar=$]
for(i=0; i<PS; i+=TS)
  #pragma [omp/oss] task \
  depend(inout: i)
  for(j=i; j<i+TS; j++) 
    do_work(j);
\end{lstlisting}
\end{minipage}
\vspace{-0.30cm}

\noindent\begin{minipage}{.45\columnwidth}
\begin{lstlisting}[caption=OMP\_TTL,label={lst:omp_t+tl},escapechar=$]
for(i=0; i<PS; i+=TS)
  #pragma omp task \
  depend(inout: i)
  {
    #pragma omp taskloop \
    grainsize(cs)
    for(j=i; j<i+TS; j++)
      do_work(j);
  }
\end{lstlisting}
\end{minipage}\hfill
\begin{minipage}{0.45\columnwidth}
\begin{lstlisting}[caption=OMP\_TF{(N)},label={lst:omp_t+f},escapechar=$]
for(i=0; i<PS; i+=TS)
  #pragma omp task \
  depend(inout: i)
  {
    #pragma omp parallel \
    for schedule(guided,cs)
    for(j=i; j<i+TS; j++)
      do_work(j);
  }
\end{lstlisting}
\end{minipage}\hfill
\vspace{-0.30cm}

\noindent\begin{minipage}{.96\columnwidth}
\begin{lstlisting}[caption=OSS\_TF{(N)},label={lst:oss_tf},escapechar=$]
for(i=0; i<PS; i+=TS)
  #pragma oss task for chunksize(cs) inout(i)
  for(j=i; j<i+TS; j++) 
    do_work(j);
\end{lstlisting}
\end{minipage}

For each of them, we have developed six different versions. \highlightForReview{
Code~\ref{lst:omp_for} implements a version using OpenMP parallel for with the 
static (OMP\_F(S)), dynamic (OMP\_F(D)) or guided (OMP\_F(G)) scheduler. 
}
Code~\ref{lst:task} shows a version using tasks in both OpenMP and OmpSs-2. This 
is a blocked version where each task computes a block of \textit{TS} size. 
Code~\ref{lst:omp_t+tl} is a version using the OpenMP taskloop. However, as 
taskloops do not accept data dependences, there is a first decomposition using 
tasks with data dependences. Then, inside each task, the block of TS size is 
partitioned anew using a taskloop. Code~\ref{lst:omp_t+f} is quite similar to 
the previous code just replacing the taskloop inside the tasks by a parallel for 
with guided scheduling, to make it similar to our worksharing tasks. Finally, 
Code~\ref{lst:oss_tf} illustrates an implementation done with worksharing 
tasks.

The N in codes~\ref{lst:omp_t+f} and~\ref{lst:oss_tf} indicates the number of 
threads used in each worksharing construct and the maximum number of 
collaborators in a worksharing task, respectively.

As a final remark, all the OpenMP implementations have been compiled and run 
with Intel OpenMP.

\subsection{Granularity evaluation in many-core architecture}
\label{subsec:granularity}
This subsection is devoted to performing a deep evaluation of our proposal in a 
many-core architecture such as the Intel KNL. So, the experiments were 
conducted on the second platform. 

In this section, we analyze the behavior of a compute-bound benchmark, the 
N-body simulation; and a memory-bound benchmark, the Stream benchmark.

In this experiment, we wish to show how the traditional ways of exploiting 
parallelism---worksharings and tasks---suffer from a lack of parallelism when 
the granularity is coarse. In that scenario, the versions using nested levels 
of parallelism---OSS\_TF, OMP\_TTL and OMP\_TF---can perform better 
because they allow higher resource utilization.

The results presented were obtained by averaging the execution times of 5 
different runs per version.

\subsubsection{\textbf{N-body simulation}}

\begin{figure}
\centering
\includegraphics[width=\columnwidth]{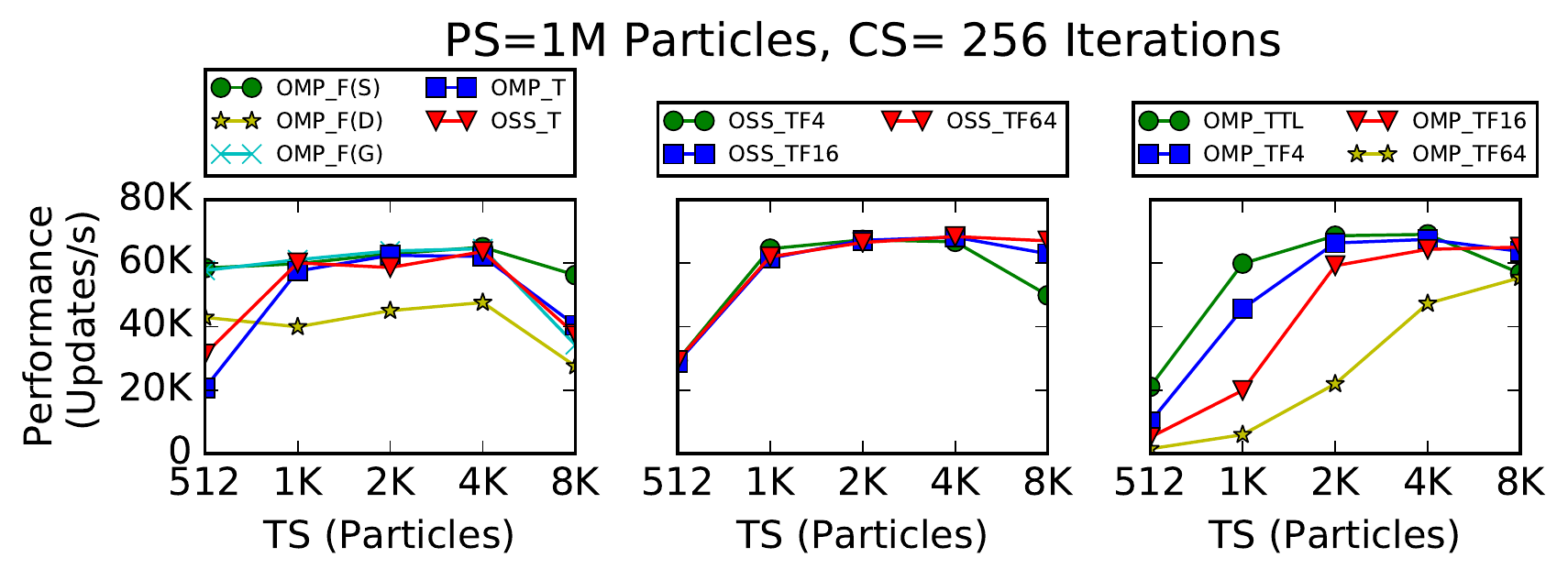}
\vspace{-0.70cm}
\caption{Granularity chart of different implementations of the N-body simulation.}
\vspace{-0.50cm}
\label{fig:nbody_granularity}
\end{figure}

Left chart of Figure~\ref{fig:nbody_granularity} compares how 
\texttt{OMP\_T}, several \texttt{OMP\_F}, and \texttt{OSS\_T} versions perform 
with different granularities. The x-axis determines the size of the blocks. The 
y-axis represents performance. For the \texttt{OMP\_F} version, TS means the 
chunksize specified in the \texttt{schedule(static,TS)} clause.

\highlightForReview{
\texttt{OMP\_F(S)} and \texttt{OMP\_F(G)} implementations perform well almost 
across the whole set of granularities but the last one. This happens because the 
heaviest computational loop contains as many iterations as the number of blocks. 
Thus, when the block size is 8K, there is a lack of parallelism because there is 
only work for ($1M/8K=128$) threads, so the other 128 are idle, and performance 
falls. Note that these versions performs well even when using quite small block 
sizes, where tasks suffer. This is because worksharing constructs introduce a 
few overhead in comparison with tasks. Notwithstanding, dynamic scheduler is 
performing quite badly. The overhead is introduced by the dynamic handling of 
chunks. It also happens with guided scheduling, but dynamic uses exactly the 
chunksize set by the user, while guided uses it as a minimum, and so may get 
bigger chunks, reducing the overall number of chunks and consequently the 
overhead.
}

The \texttt{OMP\_T} and \texttt{OSS\_T} versions start far from the worksharing 
because of the overhead introduced by tasks where the granularity is too fine. 
Then, they get peak performance until the last granularity when the performance 
falls for the same exact reason than \texttt{OMP\_F}: there is not enough 
parallelism.

The second and third chart of Figure~\ref{fig:nbody_granularity} exhibit the 
results for \texttt{OSS\_TF(N)}, and \texttt{OMP\_TF(N)} and \texttt{OMP\_TTL},
respectively. There, one can see an important difference with respect to the 
previous versions. The difference is that performance does not fall for the 
biggest granularity when N is big enough. This means that these implementations 
are able to prevent the lack of parallelism when the granularities are too 
coarse. As all these implementations are using a nested level of parallelism, 
the lack of parallelism in the outer level is alleviated by using the idle 
resources in the inner level. Consequently, peak performance is maintained for a 
broader set of granularities than traditional implementations do.

That being said, there are some other interesting points in the second and third 
charts of Figure~\ref{fig:nbody_granularity}. Firstly, it is possible to observe 
divergences among the distinct OpenMP series in the third chart. 
\texttt{OMP\_TTL} adds no extra overhead compared with using only tasks, in 
the lower granularities. Then, it gives a small boost to the performance 
in the peak granularities. Finally, for the coarser granularities, it starts 
falling, but the drop is less pronounced than the drop in the \texttt{OMP\_T} 
version. In contrast, all the \texttt{OMP\_TF} versions are introducing extra 
overhead comparing with \texttt{OMP\_T}. Note that this extra overhead becomes 
bigger as N grows. However, they are also able to provide a small increase in 
the peak, like \texttt{OMP\_TTL}.

On the other hand, the \texttt{OSS\_TF} versions, shown in the second chart of 
Figure~\ref{fig:nbody_granularity}, are not introducing further overhead with 
respect to \texttt{OSS\_T}, even with the biggest N, while they are also 
introducing a small improvement in the peak performance.

\subsubsection{\textbf{Stream benchmark}}

\begin{figure}
\centering
\includegraphics[width=\columnwidth]{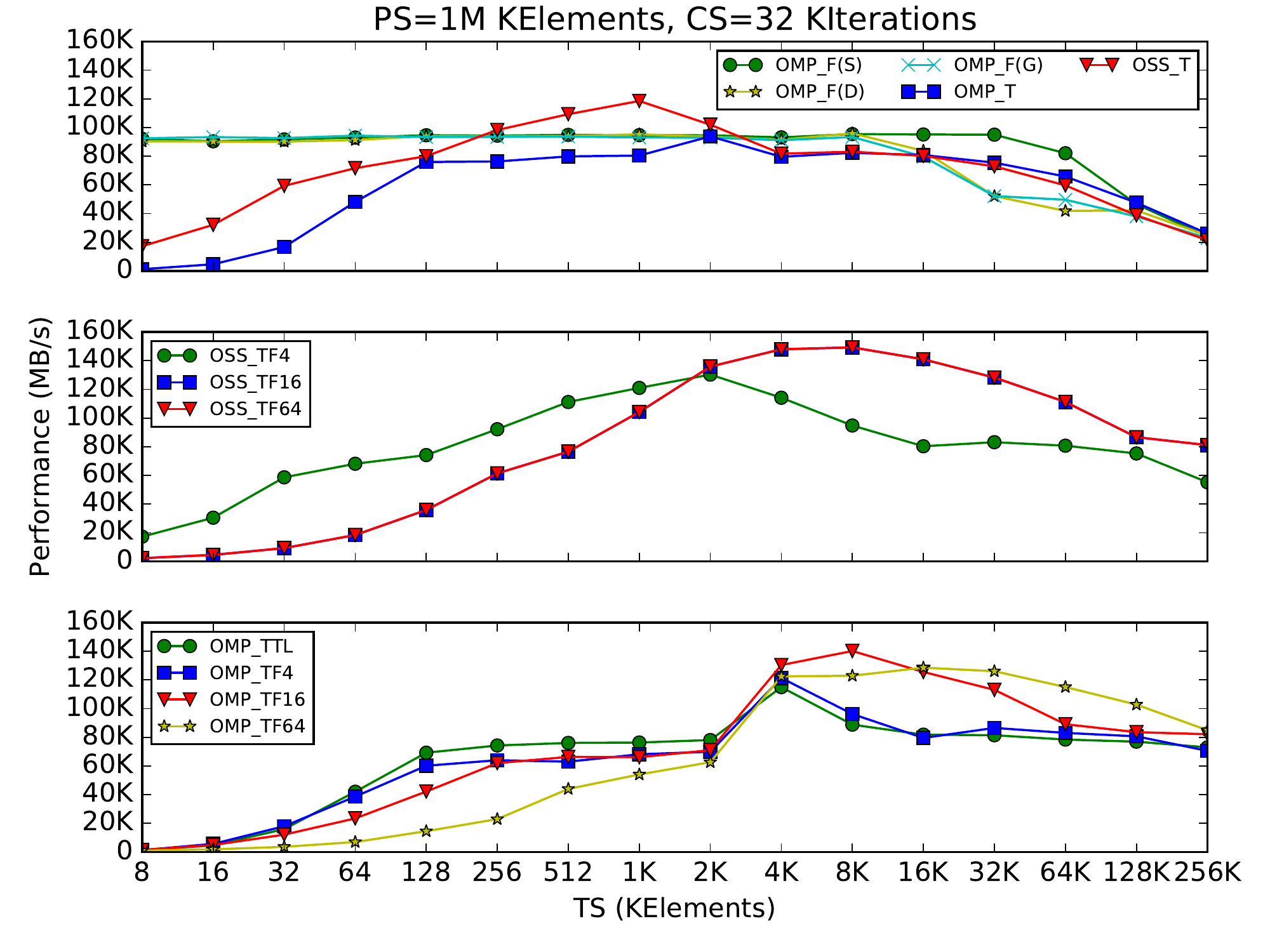}
\vspace{-0.70cm}
\caption{Granularity chart of different implementations of the Stream benchmark.}
\vspace{-0.50cm}
\label{fig:stream_granularity}
\end{figure}

The topmost chart of Figure~\ref{fig:stream_granularity} compares the 
performance of \texttt{OMP\_T}, \texttt{OMP\_F}, and \texttt{OSS\_T} versions 
using different granularities. The x-axis determines the size of the blocks. The 
y-axis represents memory bandwidth. For the \texttt{OMP\_F} version, TS means 
the chunksize specified in the \texttt{schedule(static,TS)} clause.

The main difference between the Stream benchmark and the N-body simulation is 
the weight of the computation, which is much lighter for the Stream benchmark. 
Therefore, it needs bigger granularities to hide the overhead of tasks. In the
topmost chart of Figure~\ref{fig:stream_granularity}, the first granularity in 
which tasks get good performance is 3072 Kbytes, while for the N-body 
simulation, it was 44 Kbytes, almost 70x more. Given that we need bigger 
granularities, it is more likely to end up in a granularity that constraints 
parallelism.

Looking at the topmost chart of Figure~\ref{fig:stream_granularity}, it is 
interesting to point out that the \texttt{OMP\_F} versions performs well even 
for the lowest granularity, as happened in the N-body simulation, confirming 
that it introduces very small overhead. All the \texttt{OMP\_F} versions perform 
very similarly except for the biggest granularities. The reason for that remains 
in a small change in the source code: the version with static scheduler can use 
the \texttt{nowait} clause. Hence, the \texttt{OMP\_F(S)} version only waits 
once at the end of the four loops while the rest waits four times, one per each 
loop. So, when granularities are fine, each thread runs several chunks and load 
balancing problems can be solved by the dynamic/guided scheduler. However, with
the biggest granularities, very few chunks, or even a single one, are run by 
each thread, so load balance problems make a significant difference, happening 4 
times against only 1.

Apart from this, in the same chart is possible to see the tasking versions 
outperforming all \texttt{OMP\_F} versions at some specific granularities. The 
reason for this is that tasking versions are able to exploit some data locality 
due to the immediate successor mechanism of the scheduler. With this mechanism, 
when a task finishes, if some other successor task becomes ready due to the data 
dependences release of the finished task, the successor is bypassed to the same 
CPU to exploit data locality. Finally, for the biggest granularities, there is a 
performance drop in all the versions since there is insufficient parallelism 
given that few tasks are created. For instance, for the biggest granularity, 
only ($1M/256K)=4$) tasks are created, so 252 threads are idle.

The center chart and the bottom chart of Figure~\ref{fig:stream_granularity} 
exhibit the results for \texttt{OSS\_TF(N)}, and \texttt{OMP\_TF(N)} and 
\texttt{OMP\_TTL}, respectively. Again, like for the N-body case, there exist 
important dissimilarities comparing these versions with the ones in the topmost 
chart of Figure~\ref{fig:stream_granularity}. The main one is that the biggest 
granularities are not falling so much. Again, the reason for this is that the 
additional level of parallelism introduced in these implementations palliates 
the lack of parallelism in the outer level. So, we end up having a wider set of 
granularities reaching good performance.

Interestingly, in the versions shown in the second and third chart of 
Figure~\ref{fig:stream_granularity} there is a considerable speedup in comparison 
with its tasking counterparts. For OmpSs-2, second chart, \texttt{OSS\_TF64} 
gets a 1.25x against \texttt{OSS\_T}. For OpenMP, the third chart, 
\texttt{OMP\_TF16} gets a 1.5x speedup against \texttt{OMP\_T}. The reason for 
this is that they are able to better exploit the memory hierarchy. For instance, 
when the block size is 8 KElements, using N=16, there are at most 16 tasks 
running concurrently, that means 3GB. In contrast, tasks imply N=1, so that 
means 48GB. The high bandwidth memory of the KNL, which is acting as an L3, has 
16GB of capacity. Then, for N=16, the whole dataset fits in L3, while for N=1, 
it does not.

Unlike with the N-body case, the \texttt{OSS\_TF(N)} versions, shown in the 
second chart of Figure~\ref{fig:stream_granularity}, do introduce further 
overhead with respect to \texttt{OSS\_T} when N starts growing. The same happens 
with their OpenMP counterparts, shown in the third chart, and in fact, 
\texttt{OMP\_TF(N)} versions are introducing much more overhead than 
\texttt{OSS\_TF(N)}. Anew, \texttt{OMP\_TTL} does not introduce extra overhead 
comparing to \texttt{OMP\_T}.

Overall, we would like to highlight how the set of granularities achieving peak 
performance becomes wider as N increases. This is a consequence of better 
resource exploitation. When N is small and TS is big, few tasks are created. If 
the number of created tasks is smaller than the number of concurrent teams, it 
is guaranteed that several resources will do nothing because some teams never 
get a task, hindering performance. When N grows, there are fewer teams---with 
many more collaborators---, and so it becomes more difficult for a team to get 
no tasks. Thus, it is unlikely that any of the resources remain idle. However, 
as the team size increases, the contention inside it also increases and may 
threaten performance. Overall, it looks like the best option is to use a big N, 
but still allow several concurrent teams. 

\highlightForReview{
The goal of this experiment is to show that traditional ways of exploiting 
parallelism suffer from a lack of parallelism when using coarse granularities. 
We did show that lack of parallelism on traditional approaches. However 
worksharing tasks still perform well in scenarios where traditional approaches 
do not. We can conclude that worksharing tasks offer a wider range of 
granularities delivering good performance making granularity choice easier and 
not so critical, especially when using large teams.
}
\highlightForReview{
From this experiment, we can also conclude that the number of collaborators (N) 
is important for achieving good performance. Users must take into account 
several considerations for choosing it. The first one is the \textbf{number of 
worksharing tasks}. As happens with regular tasks, the best performance is 
achieved when all the resources are busy. Therefore, if there are many tasks, N 
can be lower, since the teams will be still busy. Oppositely, if there are only 
a few tasks, N must be bigger, so that the total number of teams is reduced, and 
they can be occupied with such a low number of tasks. The second one is 
\textbf{lock contention}. Each team contains a lock which is shared among all 
the collaborators. Although it is optimized, more collaborators introduce more 
contention into the lock. Thus, using a single group with all the available cores 
may result in performance degradation. The last one is \textbf{hardware layout}. 
We do not recommend going beyond a CPU socket when setting up teams. As a 
general recommendation, we suggest using one or two teams per socket. In 
fact, the default value of our implementation is one team per socket.
}

\vspace{-0.10cm}
\subsection{Chunksize granularity}
\label{subsec:cs-granularity}
The objective of this experiment is to show that the chunksize may affect the 
performance as much as the task granularity. Thus, it must be considered and 
adequately tuned.

\begin{figure}
\centering
\includegraphics[width=\linewidth]{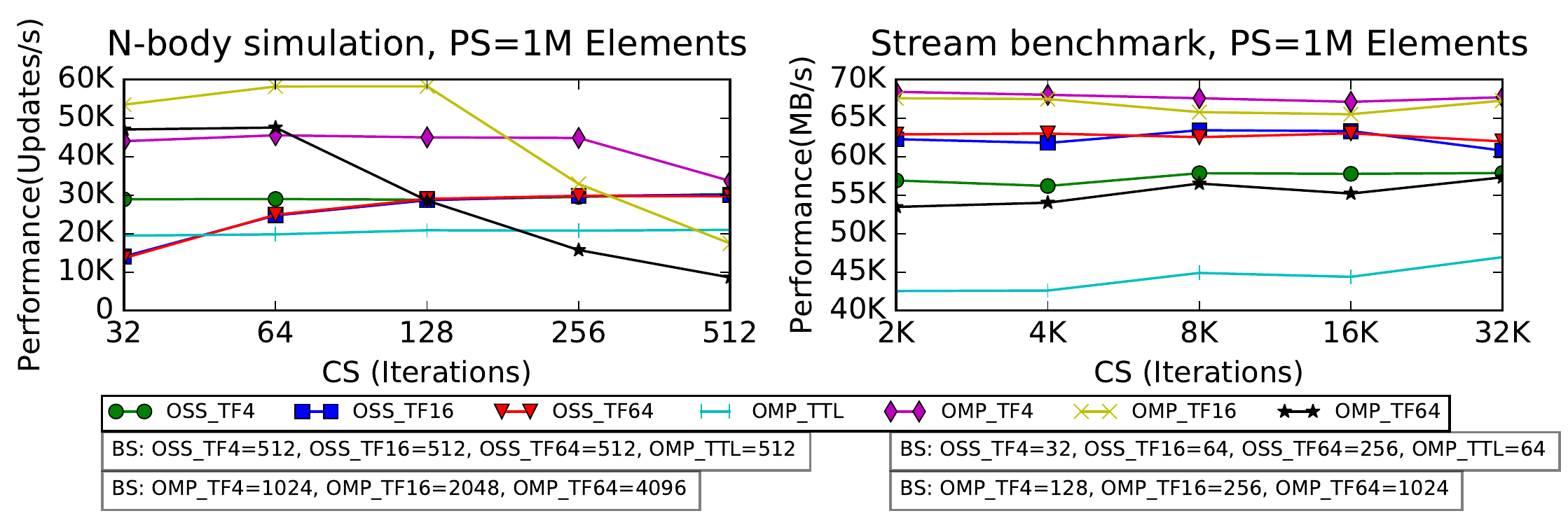}
\caption{Chunksize granularity of an N-body simulation and the Stream 
    benchmark.}
\vspace{-0.50cm}
\label{fig:cs_granularity}
\end{figure}

Figure~\ref{fig:cs_granularity} show an analysis for the N-body simulation and 
the Stream benchmark of different chunksizes for a fixed problem size. The block 
size may vary for different versions, but it is the same for all the different 
chunksizes of a version. \highlightForReview{In consequence, series cannot be 
compared against others, only points within the same series can be compared.} 
The block size is a point in the first phase of a typical granularity chart, 
where tasking overheads still hinder performance.

In the charts, the x-axis stands for chunksize (CS) in number of iterations, 
while the y-axis shows the performance for the N-body simulation, and the memory 
bandwidth for the Stream benchmark. The chunksize must be lower or equal than 
the TS, because a single chunk cannot do more iterations than those in the whole 
block.

Looking at the right chart of Figure~\ref{fig:cs_granularity}, it can be seen 
that the chunksize does not have any effect on the Stream benchmark. This 
happens because the limiting resource in this benchmark is the memory. Thus, 
waiting to acquire some locks, or letting some resources become idle, wasting 
CPU time, is not so critical like in compute-bound benchmarks. 

The chunksize is crucial in the N-body simulation, as can be seen in the left 
chart of Figure~\ref{fig:cs_granularity}. Regarding worksharing tasks, for 
medium and large values of N, an adequate chunksize provides +2x of speedup 
compared with a bad chunksize. The reason for this is that an excessively small 
chunksize may imply many more requests to the scheduler, augmenting the 
contention on the scheduler locks. Making it too large does not affect, because 
even if there are not enough chunks in a task for to feed all the workers, new 
work can start.

\highlightForReview{
In contrast, the left chart of Figure~\ref{fig:cs_granularity} shows the 
opposite behavior for OpenMP. It almost does not matter how small the chunksize 
is. The OpenMP guided policy assigns chunks dynamically. The actual chunk size 
is proportional to the number of unassigned chunks divided by the number of 
threads in the team, with the costraint that it can never be lower than the 
value set by the user. So, usually, big chunks are assigned at the beginning. 
Then, they become smaller and smaller until the last iterations where the 
restriction appears. So, when a user sets a chunksize too small, it only affects 
a few chunks at the end of the execution, and so it does not make a big 
difference. However, it is affected if the chunksize becomes too big since it 
cannot feed all the cores and some of them may wait in the barrier until the 
rest finish.
}

\highlightForReview{
We have evidenced that chunksize may be important in some applications; while 
completely nimium in others. As a general recommendation, we suggest using 
$CS=TS/N$ so that each collaborator in the team has at least one chunk to 
execute. Nonetheless, having at least one chunk per collaborator is not really 
important if we have several ready tasks at the same time, because in that case, 
collaborators can get new tasks. In contrast, when there are only a few ready 
tasks, it is important to have as many chunks as collaborators or they will 
probably remain idle. Furthermore, the cost of the computation is also 
important. Heavier computations can work well with lower chunksizes while 
lighter computations will require bigger chunksizes to palliate the overheads.
}

\subsection{Strong scaling}
\label{subsec:strong-scaling}
This experiment consists in fixing a given number of resources and decreasing 
the problem size, obtaining a smaller problem size per core at each new point 
of the experiment. The goal is to illustrate that there exist scenarios where 
the problem size per core prevents the possibility of setting an adequate 
granularity. In these scenarios, either task management overheads---if the 
granularity is too fine--- or lack of parallelism---if the granularity is too
coarse--- hinders performance. Thus, by using nested levels of parallelism that allow 
the use of coarse-grained tasks that are then split into several chunks, 
performance improves. For this experiment, we have used all the 
benchmarks presented in Section~\ref{subsec:benchmarks}.

The results of the experiment are presented in two charts per benchmark, one per 
platform. In these charts, in the x-axis, there are different problem sizes. The 
left y-axis represents performance while the right y-axis stands for work units 
per hardware thread. The charts show four different series (bars) for each 
problem size. \highlightForReview{Those series are six different implementations, \texttt{OMP\_F(S)},
\texttt{OMP\_F(D)}, \texttt{OMP\_F(G)}, \texttt{OMP\_T}, \texttt{OSS\_TF} and, 
finally, the one obtaining best performance between \texttt{OMP\_TF(N)} and 
\texttt{OMP\_TTL}.} For each of the bars, there is also a circle pointing out 
the number of work units per hardware thread for that specific configuration. 
Finally, there is a horizontal line which corresponds to 1 work unit per 
hardware thread. Thus, it is easy to see when there is at least work for all the 
resources (above the line) and when there is not (below the line).

\highlightForReview{
For all the versions, we have explored the whole set of combinations for each 
of the parameters (TS, CS and N if applicable), and selected the best 
configuration.
}

\begin{figure}
\centering
\includegraphics[width=\linewidth]{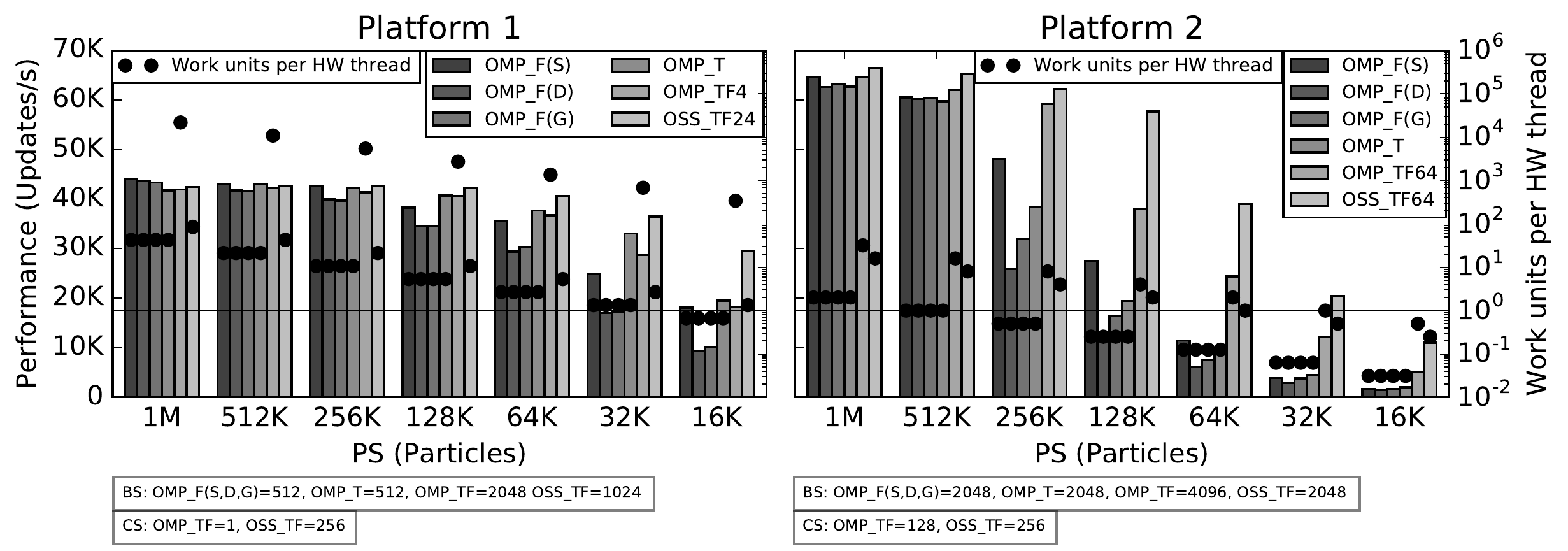}
\vspace{-0.70cm}
\caption{Strong scaling charts of the N-body simulation.}
\vspace{-0.50cm}
\label{fig:strong_scaling_nbody}
\end{figure}

\highlightForReview{
Figure~\ref{fig:strong_scaling_nbody} shows the results for the N-body 
simulation. In the first platform, left chart of 
Figure~\ref{fig:strong_scaling_nbody}, all the implementations perform very 
similarly for the three biggest problem sizes, with all \texttt{OMP\_F} versions 
standing out a bit for the biggest problem size. Then, performance decreases for 
all the versions. For the biggest problem sizes, all \texttt{OMP\_F} versions 
deliver similar performance. Then, \texttt{OMP\_F(S)} outperforms its dynamic 
and guided counterparts. Dynamic and guided schedulers introduce more overhead 
than static. They are worth if the application is highly imbalanced, but this is 
not the case. Hence, they are introducing overhead but not getting any benefit, 
hurting performance. It is not significant in the biggest problem sizes because 
the long execution time amortizes the overhead, but it pops up when the 
execution shortens.
}

Regarding \texttt{OMP\_F} and \texttt{OMP\_T} versions, except for the lowest 
problem size, there is at least one work unit for each hardware thread, so only 
the lowest granularity has a lack of parallelism. From 32K to 128K, there is a 
load balancing problem. There are, respectively, 5.33, 2.67, and 1.33 work units 
per thread for 128K, 64K, and 32K. This means that some threads are performing 
more work than others, and those others are just idle wasting resources.

\texttt{OMP\_TF} version has more than enough parallelism when considering both 
levels of parallelism, but the nested parallel regions are introducing a lot of 
overhead, and that hurts performance. Additionally, for the lowest problem size, 
there is not enough parallelism in the first level to feed all the resources, so 
that even having enough work units when considering combined parallelism, those 
work units are concentrated in only half of the resources, remaining the rest 
idle.

In contrast, it can be seen how \texttt{OSS\_TF} is able to maintain the 
performance much better than the other versions, reaching up to a 1.5x speedup 
for the lowest size against the best competitor. Note that the problem size is 
reduced by up to 64x, but \texttt{OSS\_TF} performance is still 70\% of the 
peak performance while the rest are below 50\%. The main reason is that even 
for the lowest size, we reach high levels of hardware resources ocuppancy thanks 
to having very few (concretely 2) teams with a high amount of CPUs. This allows 
not only the parallelism to be maximized but also to improve load balancing.

In the second platform, the right chart of Figure~\ref{fig:strong_scaling_nbody}, 
the behavior is very similar but accentuated because of the large number of 
cores available. The performance of \texttt{OMP\_F} and \texttt{OMP\_T} falls 
very quickly because of the lack of parallelism. In contrast, both 
\texttt{OMP\_TF} and \texttt{OSS\_TF} are able to maintain acceptable 
performance even when there is not enough parallelism in the first level, thanks 
to its nested level of parallelism. Nevertheless, \texttt{OSS\_TF} outperforms 
its OpenMP equivalent, becoming the difference between them bigger as the 
problem size per core decreases. \texttt{OSS\_TF} is able to get up to 2x 
speedup compared with \texttt{OMP\_TF} and more than 5x compared with 
\texttt{OMP\_T} and \texttt{OMP\_F}.

\begin{figure}
\centering
\includegraphics[width=\linewidth]{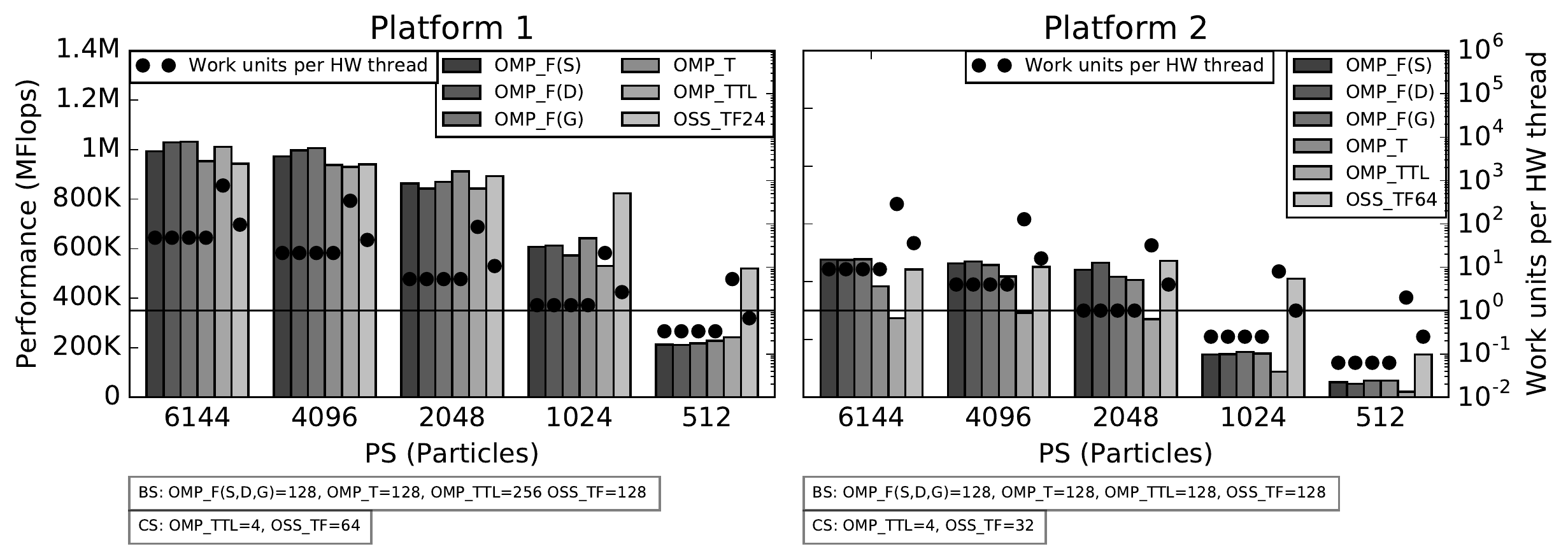}
\vspace{-0.70cm}
\caption{Strong scaling charts of the matmul benchmark.}
\vspace{-0.50cm}
\label{fig:strong_scaling_matmul}
\end{figure}

Figure~\ref{fig:strong_scaling_matmul} shows the results for the MATMUL 
benchmark.

Regarding the results of the first platform, displayed in the left chart of 
Figure~\ref{fig:strong_scaling_matmul} it is possible to observe one more time 
the performance reduction as the problem size becomes smaller. The reasons are 
load balancing, like for the N-body simulation, for PS=1024; and the lack of 
parallelism for PS=512. Yet, \texttt{OSS\_TF} keeps performance better than the 
other versions, achieving up to a 2x speedup against the best OpenMP version.

The nature of the chart of the second platform, shown in the right chart of 
Figure~\ref{fig:strong_scaling_matmul}, is similar to the previous, but in 
this case, the main problem is actually the lack of parallelism given the large 
number of available resources. In this platform, nonetheless, \texttt{OSS\_TF} 
is able to reach up to a 2.7x speedup versus the best OpenMP version.

\begin{figure}
\centering
\includegraphics[width=\linewidth]{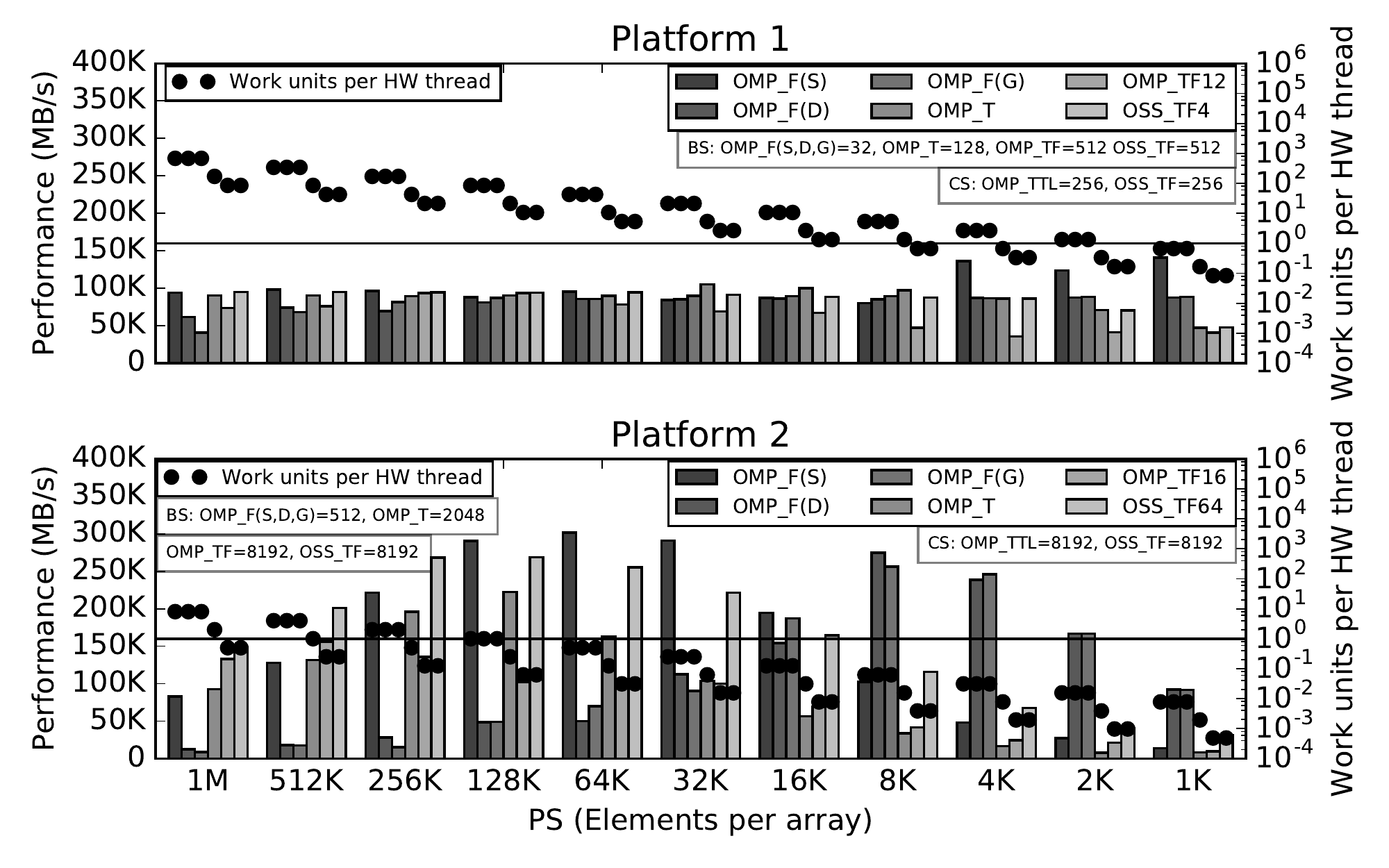}
\vspace{-0.70cm}
\caption{Strong scaling charts of the Stream benchmark.}
\vspace{-0.50cm}
\label{fig:strong_scaling_stream}
\end{figure}

The results of the Stream benchmark, available in 
Figure~\ref{fig:strong_scaling_stream}, are different than the previous. 
In the previous benchmarks, there was a trend where lowering the problem size 
led to a performance drop, especially in the OpenMP versions.

In the first platform, the topmost chart of Figure~\ref{fig:strong_scaling_stream}, 
this does not happen, or at least, the drop is not as large. The main reason is 
that the limiting resource in this benchmark is the memory bandwidth instead of 
the CPU. Thus, even without using all the resources, peak performance can be 
achieved. For this reason, decreasing the problem size, leading to a lack of 
parallelism, is not so important in this benchmark.

\highlightForReview{
That being said, we can see how the \texttt{OMP\_F} versions are even increasing 
its performance, especially \texttt{OMP\_F(S)}. The increase stems from the data 
locality exploitation given that in the lowest sizes, the whole problem or a 
large part of it fits in the caches. This effect is seen in none of the other 
versions mainly due to two reasons. The first one is the pollution of the caches 
caused by the runtime libraries. The second one is that the \texttt{static} 
scheduling of the \texttt{OMP\_F(S)}, combined with the nowait clause, allows 
that a CPU executes the same elements of each loop, maximizing data reuse. In 
contrast, task-based versions, although they have immediate successor policy 
which favors locality, is not so perfect as the \texttt{OMP\_F(S)} one. 
Regarding the dynamic and guided versions, they need to run the whole loop, 
iterating over the whole data arrays before moving forward to the next loop, 
preventing them from any kind of data reuse. 
}

In the second platform, the bottom chart of Figure~\ref{fig:strong_scaling_stream}, 
the effect of insufficient parallelism is notable, like in the previous 
benchmarks. It is caused by the large number of cores available in this 
platform, which needs a bigger value of problem size to keep the problem 
size per core able to perform decently. In this platform, there are four 
versions that stand out. Again, the reason is data locality. This platform 
incorporates a 16GB high bandwidth memory used as L3. The problem sizes where we 
get the best performance are those where the whole data set fits in cache while 
there is enough parallelism. 

\begin{figure}
\centering
\includegraphics[width=\linewidth]{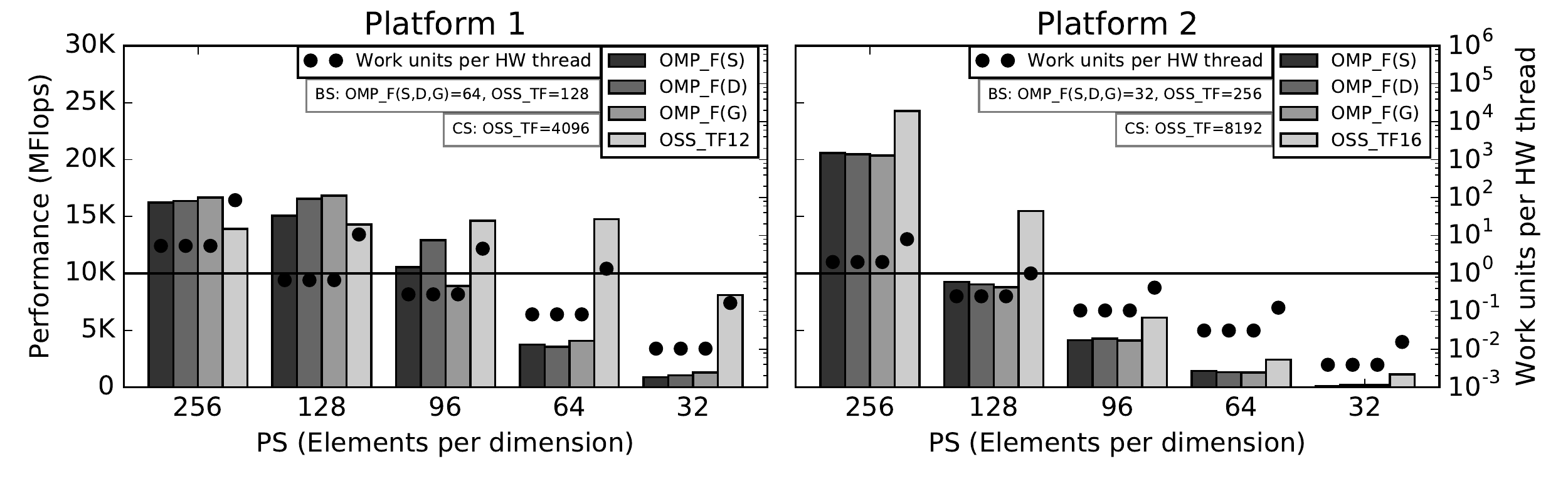}
\vspace{-0.70cm}
\caption{Strong scaling charts of the HPCCG benchmark.}
\vspace{-0.50cm}
\label{fig:strong_scaling_hpccg}
\end{figure}

Figure~\ref{fig:strong_scaling_hpccg} introduces the results obtained in the 
HPCCG benchmark. In this benchmark, each chart only contains two 
series instead of the four mentioned previously. The reason is that the HPCCG 
benchmark contains several reductions. The Intel OpenMP compiler and runtime 
we have used do not support task reductions. Therefore, all the versions using 
tasks (\texttt{OMP\_T}, \texttt{OMP\_TF} and \texttt{OMP\_TTL}) perform 
poorly. 

The results of the first platform are given in the left chart of 
Figure~\ref{fig:strong_scaling_hpccg}. In that chart, it is possible to see 
again, like in the previous benchmark, how the performance of all \texttt{OMP\_F} 
versions drops when the problem size per core decreases. In contrast, the 
performance of \texttt{OSS\_TF} remains very similar until the lowest problem 
size where it finally drops. Despite the drop, \texttt{OSS\_TF} gets more than 
9x speedup compared to \texttt{OMP\_F} for that problem size. 
The reason for the drop, one more time, is the lack of parallelism. However, it 
is possible to see some of the circles below the line while the performance is 
still peak. This happens for the same reason than the Stream benchmark. HPCCG is 
also memory-bound, so it does not need to occupy all the cores to reach 
peak performance. \highlightForReview{There are also some differences depending 
on the scheduler for the \texttt{OMP\_F} versions. Dynamic and guided seems to 
perform slightly better. The reason is load imbalance.}

In the second platform, the right chart of Figure~\ref{fig:strong_scaling_hpccg}, 
the trends are similar. \texttt{OMP\_F} performance is deteriorated by the lack 
of parallelism. However, the \texttt{OSS\_TF} performance in this platform, 
falls faster because there are many more resources available, and even 
\texttt{OSS\_TF} is not able to exploit enough of them when the problem size is 
reduced. Still, \texttt{OSS\_TF} outperforms \texttt{OMP\_F} by up to 1.65x.

\highlightForReview{
We have demonstrated that when performing strong scaling experiments we can 
easily get into scenarios where the problem size per core prevents traditional 
ways of exploiting parallelism to get good performance. At the same time, we 
have shown how worksharing tasks mitigate the lack of parallelism issue being 
able to perform well across several benchmarks on two different platforms, even 
in scenarios where traditional approaches suffer.
}
  
\vspace{-0.2cm}
\section{Conclusions}
\label{sec:conclusions}
In this paper, we propose a new concept called worksharing tasks that leverage 
the flexibility of tasks and the efficiency of worksharing techniques to exploit 
fine-grained loop parallelism. Our proposal introduces the new \texttt{for} 
clause---\texttt{do} clause in Fortran---to the task construct. 

A worksharing task is like a regular task that encompasses a for loop. The key 
difference is that this for loop can be run by several workers using worksharing
techniques that have been adapted to avoid any fork-join synchronization 
to preserve the fine-grained data-flow execution model of regular tasks. 

In general, task-based programming models require at least one task per core to
achieve the best performance. This fact provides an upper bound on the task 
granularity, which proportionally increases with the problem size but 
proportionally decreases with the number of cores. Thus, a small problem size 
combined with a large number of cores limits task granularity and impacts
performance.  

Using too fine-grained tasks, the overheads related to task management hinder 
performance; while using too coarse-grained tasks, the number of tasks is not 
enough to fully exploit all cores. The lower bound of task granularity that 
reaches peak performance is determined by the efficiency of the runtime system 
to handle tasks, while the upper bound of task granularity is determined by the 
problem size per core. 

Worksharing tasks overcome the requirement of one task per core to
achieve high resource utilization by allowing a small number of coarse-grained 
worksharing tasks that are partitioned into several fine-grained chunks.
Worksharing tasks allow us to increase task granularity up to 64x without 
limiting the available parallelism. Hence, a small number of worksharing tasks 
can efficiently exploit a many-core processor. 

Our evaluation shows that worksharing tasks not only outperform traditional 
tasks and worksharing techniques, but also advanced combinations of both 
techniques. Worksharing tasks get up to 9x speedup against the most performant 
OpenMP alternative in some scenarios. Moreover, the use of worksharing tasks 
increases the range of granularities that reach peak performance. 
Finally, our proposal does not add any coding complexity over the traditional 
task-based implementation.

\section{Future work} 
\label{sec:future}
We plan to investigate additional scheduling policies to distribute loop 
iterations across workers, as well as dynamic composition of the teams that 
execute a worksharing task, with the goal of improving the flexibility and 
efficiency of this mechanism. 

Also, we intend to explore the interaction between the OpenMP taskloop   
and our worksharing tasks. Currently, taskloop distributes a loop into tasks. 
Given that worksharing tasks are essentially tasks, this should naturally work.


\section*{Acknowledgment}
This work is supported by the Spanish Ministerio de Ciencia, Innovaci\'{o}n y
Universidades (TIN2015-65316-P), by the Generalitat de Catalunya
(2014-SGR-1051) and by the European Union's Seventh Framework Programme 
(FP7/2007-2013) and the H2020 funding framework under grant 
agreement no. H2020-FETHPC-754304 (DEEP-EST).

\bibliographystyle{IEEEtran}
\vspace{-0.2cm}
\bibliography{bib/main.bib}


\end{document}